 \documentclass[prb,twocolumn,showpacs,superscriptaddress]{revtex4-1}

 \usepackage{amsmath} 
  \usepackage{graphicx,verbatim,mathenv}
 
\def\beq {\begin{equation}}
\def\eeq {\end{equation}}

\def\beqn  {\begin{displaymath}}
\def\eeqn {\end{displaymath}}

\def\beqarr {\begin{multline}}
\def\eeqarr {\end{multline}}

\def\bv {{\mid}}

\def\w {\omega}

\def\bfr {\mathbf{r}}
\def\bfq {\mathbf{q}}

\def\bfG {\mathbf{G}}

\def\contra {\sqcap}

\begin{document}
  \title{Design of effective kernels for spectroscopy and molecular transport: time-dependent current-density-functional theory}
  \author{Matteo Gatti}
\altaffiliation[Permanent address: ]
{Centro Joxe Mari Korta, Avenida de Tolosa, 72, E-20018 San Sebasti\'an, Spain.}
   \affiliation{Nano-Bio Spectroscopy group, 
Dpto. F\'isica de Materiales, Universidad del Pa\'is Vasco, 
Centro de F\'isica de Materiales CSIC-UPV/EHU-MPC and DIPC, 
E-20018 San Sebasti\'an, Spain}
 \affiliation{European Theoretical Spectroscopy Facility (ETSF)}
\affiliation{Laboratoire des Solides Irradi\'es, \'Ecole Polytechnique, CNRS-CEA/DSM,  F-91128 Palaiseau, France}
 \date{\today}

\begin{abstract}
Time-dependent current-density-functional theory (TDCDFT) provides an in principle exact scheme to calculate efficiently response functions for a very broad range of applications. However, the lack of approximations valid for a range of parameters met in experimental conditions has so far delayed its extensive use in inhomogeneous systems. 
On the other side, in  many-body perturbation theory (MBPT) accurate approximations are available, but at a price of a higher computational cost.
In the present work the possibility of combining the advantages of both approaches is exploited.
In this way an exact equation for the exchange-correlation kernel of TDCDFT is obtained, which opens the way for a systematic improvement of the approximations adopted in practical applications. 
Finally, an approximate  kernel for an efficient calculation of spectra of solids and molecular conductances is suggested and its validity discussed.

  \end{abstract}
 
 

  \maketitle

The theoretical description of the response of an electronic system to a time-dependent perturbation is a key problem for many areas of physics and chemistry.
Most spectroscopic experiments probe the elementary excitations of an electronic system through its linear response to an external electromagnetic field.
Their theoretical interpretation is of primary interest for technological applications in condensed-matter physics, nanosciences, photochemistry or biophysics.
Similarly, response functions are essential, among many possible applications, also for the determination of the electrical conductivity or other transport coefficients in molecular electronic devices. 

Therefore, one would like to devise a reduced theoretical framework that is, at the same time, \emph{reliable} and \emph{efficient}, by calculating only the information needed to interpret and predict specific experimental measurements.
Two prominent examples of such reduced approaches are many-body perturbation theory (MBPT) and density-functional-based theories.
Key variables of the former are one- and two-particle Green's functions, $G(1,2)$ and $G_2(1,2,3,4)$ ($1$ is a shorthand notation for space, time and spin indices $\bfr_1,t_1,\sigma_1$). Methods based on the Green's-function formalism reduce the complexity of the many-body wavefunction into the propagation and the interaction of renormalized quasiparticles.
Their intuitive, direct contact with the initial problem of real interacting electrons makes it rather \emph{easy} to introduce working approximations. 
A remarkably successful example is the solution of the Bethe-Salpeter equation (BSE) for the two-particle correlation function $L(1,2,3,4)=-G_2(1,2,3,4)+G(1,3)G(2,4)$ which has led to an important breakthrough by permitting, for instance, an accurate calculation of electronic spectra of solids and nanosystems \cite{RMP-NOI}. On the other hand, practical calculations  at this level remain very demanding, even for nowadays' computers.
An alternative pathway is instead based on the extension of density-functional theory (DFT) to scalar time-dependent external potentials, $V_{ext}(\bfr,t)$,
as in time-dependent density-functional theory (TDDFT)\cite{tddft}, or to time-dependent vector potentials, $\mathbf{A}_{ext}(\bfr,t)$,
as in time-dependent current-density-functional theory (TDCDFT)\cite{Ghosh88,vignaleprl}. 
When one needs only charge or current densities, $\rho(\bfr,t)$ and $\mathbf{j}(\bfr,t)$, these density-based methods identify the minimum content of information that one has to calculate in order to provide the searched answers.
In the Kohn-Sham (KS) scheme \cite{KS}, the many-body problem is reformulated very efficiently into a set of self-consistent non-interacting one-particle equations. For this reason, in the KS scheme the solution of the full many-body problem is made \emph{simple} and its computational cost is very convenient.
The main drawback is that it is generally very difficult to improve upon the simplest local-density approximations (LDA) \cite{KS} for the exact density functionals of the formal theory. In fact, after the first promising results \cite{metasi}, the development of TDCDFT in this field has been delayed by the lack of adequate approximations to the tensor exchange-correlation (xc) kernel $\hat{f}_{xc}$,  beyond the local functional in the current density  of Vignale and Kohn (VK) \cite{vignaleprl,vignaleprl2}, which is unfortunately valid only in a range of parameters that is often not met in experiments performed on inhomogeneous systems \cite{metano,berger}.

The present work aims at overcoming the limitations of the VK functional, by applying to TDCDFT the emerging successful strategy of combination of MBPT and density-functional approaches, in order to profit from the complementary advantages of both \cite{fabien,longrange,nanoquanta,stubner,ulf}. 
In particular, we will derive an exact relation linking the unknown exchange-correlation tensor kernel $\hat{f}_{xc}$   to quantities that can be in principle  accurately calculated in MBPT.  Moreover, we will show how it is possible to introduce suitable approximations to this exact relation opening the way to a broad class of applications and to a systematic way to improve the approximations adopted in TDCDFT calculations.

The Bethe-Salpeter equation for the irreducible polarization function $\tilde{L}(1,2,3,4)$  reads\cite{Strinati} (throughout the paper, integrals and sums are always done on repeated indices and atomic units are adopted):
\begin{multline}
\tilde{L}(1,2,3,4) = L_0(1,2,3,4) + \\ L_0(1,2,5,6) \tilde{\Xi}(5,6,7,8) \tilde{L}(7,8,3,4).
\label{eqbse10} 
\end{multline}
Here $L_0(1,2,3,4) = -iG(1,3)G(4,2)$ is the two-particle correlation function for independent particles and  $\tilde{\Xi}(5,6,7,8) = i \delta\Sigma(5,6) / 
\delta G(7,8)$ is the BSE kernel, which e.g. accounts for excitonic effects in optical spectra.
In standard BSE implementations the GW approximation \cite{hedin} for the self-energy $\Sigma$ is adopted, where $\Sigma$ is evaluated as a product of the one-electron Green's function $G$ and the screened Coulomb interaction $W$. $\tilde{\Xi}$ is usually approximated as $\tilde{\Xi}(5,6,7,8) =-W(5,6)\delta(5,7)\delta(6,8)$. 
This amounts to neglecting the term $iG {\delta}W/{\delta}G$, which contains information about the change of the screening in the excitation and is considered to be small.
Moreover, for $W$ one generally considers only a static screening of the Coulomb interaction $v$ and $L_0$ is built with GW quasiparticle (QP) energies and KS wavefunctions.
$\tilde{L}$ is then linked to the correlation function $L$ by a Dyson equation:  $L=\tilde{L}+\tilde{L}vL$.
Whereas the quantities of spectroscopic interest are for instance the two-point density-density and current-current response functions, $\chi_{\rho\rho}(1,2)=\delta \rho(1) / \delta V_{ext}(2)$ and $\hat{\chi}
(1,2) =\delta \mathbf{j}(1) / \delta \mathbf{A}_{ext}(2)$,  the BSE  is an intrinsically four-point equation. In fact, in the BSE scheme, these two-point response functions can be obtained only as contractions of the four-point correlation function $L$, which has to be calculated  in a first step.
In many situations, as the ones we are interested in here, this clearly reveals to be a computational waste that one would like to avoid.

In TDCDFT the linear response of the current $\mathbf{j}$ to an external vector potential $\mathbf{A}_{ext}$ is\cite{vignaleprl}: 
\beq
\delta j_\alpha(1) =  \frac{1}{c} \hat{\chi}_{s,\alpha\beta}(1,2)\delta A_{s,\beta}(2),
\label{tdcdft1}
\eeq
where the Kohn-Sham vector potential $\mathbf{A}_s$ is the sum of the external, Hartree and exchange-correlation potentials: 
$\mathbf{A}_s(1) = \mathbf{A}_{ext}(1)+ \mathbf{A}_{H}(1)+ \mathbf{A}_{xc}(1)$, and $\hat{\chi}_{s}$ is the Kohn-Sham current-current response function.
Similarly, the linear variation of the current can be calculated through the knowledge of the irreducible current-current response function $\hat{\tilde{\chi}}$:
\beq
\delta j_\alpha(1) =  \frac{1}{c} \hat{\tilde{\chi}}_{\alpha\beta}(1,2) [\delta A_{ext,\beta}(2)+ \delta A_{H,\beta}(2) ].
\label{tdcdft2}
\eeq
Combining these two definitions, one immediately gets to a Dyson equation linking  $\hat{\chi}_{s}$ with $\hat{\tilde{\chi}}$:
\begin{multline}
\hat{\tilde{\chi}}_{\alpha\beta}(1,2) = \hat{\chi}_{s,\alpha\beta}(1,2) + \\ \hat{\chi}_{s,\alpha\lambda}(1,3)\hat{f}_{xc,\lambda\kappa}(3,4) \hat{\tilde{\chi}}_{\kappa\beta}(4,2),
\label{tdcdft3}
\end{multline}
where the exchange-correlation tensor kernel
\beq
\hat{f}_{xc,\alpha\beta}(1,2) = \frac{\delta A_{xc,\alpha}(1) }{\delta j_\beta(2) }
\label{tensorkernel}
\eeq
has been introduced. Once thanks to \eqref{tdcdft3} the irreducible $\hat{\tilde{\chi}}$ has been calculated, the (reducible) response function $\hat{\chi}$ (hence the spectra)  can be obtained through:
\begin{multline}
\hat{\chi}_{\alpha\beta}(\bfr_1,\bfr_2,\w)= \hat{\tilde{\chi}}_{\alpha\beta}(\bfr_1,\bfr_2,\w) + \\
- \hat{\tilde{\chi}}_{\alpha\lambda}(\bfr_1,\bfr_3,\w) \frac{1}{\w^2} \nabla_{3_\lambda} \frac{1}{\bv\bfr_3-\bfr_4\bv} \nabla_{4_\kappa}  \hat{\chi}_{\kappa\beta}(\bfr_4,\bfr_2,\w).
\end{multline}
The quality of the approximation adopted for the xc kernel \eqref{tensorkernel} is hence fundamental for the accuracy of the final results.

As in TDDFT, also in TDCDFT the effect of the xc kernel on the spectra calculated from the independent KS-particle response is twofold. 
The Kohn-Sham eigenvalues are known to underestimate the quasiparticle band gap of insulating systems due to the derivative discontinuity of the DFT xc potential\cite{ss83}. Hence the xc kernel has first to provide a consistent band gap opening. And, second, as the BSE kernel $\tilde{\Xi}$, it has to describe electron-hole interactions. 
So, following Refs. \cite{fabien,stubner}, in order to make explicit these two aspects, here we set $ \hat{f}_{xc,\alpha\beta}= \hat{f}_{xc,\alpha\beta}^{(1)}+ \hat{f}_{xc,\alpha\beta}^{(2)}$, where $\hat{f}_{xc,\alpha\beta}^{(1)}$ has the task of overcoming the KS band gap problem, while $\hat{f}_{xc,\alpha\beta}^{(2)}$, for instance, accounts for excitonic effects in optical spectra, or dynamical corrections to the Landauer formula for the electronic conductance in quantum transport\cite{landauer}. Formally, we will now introduce a contraction operator $\contra_{\alpha\beta}$:
\begin{multline}
\contra_{\alpha\beta} L_0(1,1',2,2') = \\
\frac{1}{2i} \frac{1}{2i} \left [  (\nabla_{1_\alpha}-\nabla_{1'_\alpha})  
(\nabla_{2_\beta}-\nabla_{2'_\beta}) L_0(1,1',2,2') \right ]_{1'=1^+,2'=2^+}, \label{chioab} 
\end{multline}
in such a way that $\hat{\chi}_0$ is\cite{Strinati}:
\beq
 \hat{\chi}_{0,\alpha\beta}(1,2) =  \rho(1)\delta(1,2)\delta_{\alpha\beta}  +  \contra_{\alpha\beta}  L_0(1,1',2,2').
\eeq 
$\hat{\chi}_0$  is built with QP ingredients instead of KS ones. 
In this way one has:
\beq
 \left[\hat{\chi}_s^{-1}\right]_{\alpha\beta}(1,2) - \left[\hat{\chi}_0^{-1}\right]_{\alpha\beta}(1,2) =\hat{f}_{xc,\alpha\beta}^{(1)}(1,2),
\eeq
and, then:
\beq
\left[\hat{\chi}_0^{-1}\right]_{\alpha\beta}(1,2) - \left[\hat{\tilde{\chi}}^{-1}\right]_{\alpha\beta}(1,2) =  \hat{f}_{xc,\alpha\beta}^{(2)}(1,2).
\label{eqtddft}
\eeq
In many semiconductors, the difference between KS and QP can be accounted for by using a scissor operator that shifts rigidly upwards the eigenvalues of the conduction with respect to valence bands. The use of a scissor operator is also a common practice in TDCDFT \cite{berger,deboeij}. The main point of interest in our discussion here is hence about $f_{xc,\alpha\beta}^{(2)}$. 

Eq. \eqref{eqtddft} leads to:
\begin{multline}
\hat{\tilde{\chi}}_{\alpha\beta}(1,2)=\hat{\chi}_{0,\alpha\beta}(1,2)+ \\
\hat{\chi}_{0,\alpha\lambda}(1,3)\hat{f}_{xc,\lambda\kappa}^{(2)}(3,4) \hat{\tilde}{\chi}_{\kappa\beta}(4,2).
\label{eqtddft2}
\end{multline}
By definition, both TDCDFT and BSE yield the exact two-point response function $\hat{\tilde{\chi}}_{\alpha\beta}(1,2)=  \rho(1)\delta(1,2)\delta_{\alpha\beta}  +  \contra_{\alpha\beta} \tilde{L}(1,1',2,2')$.
Therefore, applying the contraction $\contra_{\alpha\beta}$ to Eq. \eqref{eqbse10} and comparing the result with Eq. \eqref{eqtddft2}, one obtains:
\begin{multline}
\hat{\chi}_{0,\alpha\lambda}(1,3)\hat{f}_{xc,\lambda\kappa}^{(2)}(3,4) \hat{\tilde{\chi}}_{\kappa\beta}(4,2) = \\
\contra_{\alpha\beta} \{L_0(1,1',3,4)  \tilde{\Xi}(3,4,5,6)  \tilde{L}(5,6,2,2')\},
\label{eq_eff72}
\end{multline}
which can be solved  for the TDCDFT kernel $\hat{f}_{xc,\alpha\beta}^{(2)}$ giving:
\begin{multline}
\hat{f}_{xc,\alpha\beta}^{(2)}(1,2) = \bigl[\hat{\chi}_0^{-1}\bigr]_{\alpha\lambda}(1,3) 
\contra_{\lambda\kappa} 
\left[
 L_0(3,3',4,5) \right. \\ 
 \left. \tilde{\Xi}(4,5,6,7)  \tilde{L}(6,7,8,8') 
\right] \bigl[ \hat{\tilde{\chi}}^{-1}\bigr]_{\kappa\beta}(8,2).
\label{eq_eff73}
\end{multline}

Eq. \eqref{eq_eff72}  is a generalized Sham-Schl\"uter equation \cite{ss83,SSE}, which relates in an {\it exact} manner TDCDFT quantities with MBPT ones, opening the way to possible systematic improvements in the design of new approximations for the xc kernel of TDCDFT. 
Its main worth is that it allows avoiding to approximate directly the kernel $\hat{f}_{xc}$, which can be a difficult problem. 
The design of operative approximations is instead simpler in the context of MBPT.
Then, thanks to  Eq. \eqref{eq_eff73}, working approximations of MBPT can be mapped into the more efficient TDCDFT scheme, where one would prefer to solve the equations. In particular, when approximating Eq. \eqref{eq_eff73}, no assumptions of locality in the current density of the xc functional are explicitly needed, leading to approximate kernels that can be employed also in the range of parameters where the VK functional is formally not valid, namely
for ground-state densities and induced current densities not slowly varying in space, and in the region below the particle-hole continuum of the homogeneous electron gas. 

Within the TDDFT framework, such a mapping strategy has already demonstrated to be a successful approach and has led to the introduction of an xc kernel, known as Nanoquanta kernel \cite{fabien,longrange,nanoquanta,stubner,ulf,SSE}, which has shown to provide the same level of accuracy as the BSE in a wide range of spectroscopy applications, from solids to finite molecular chains \cite{botti,varsano}. Therefore, by discussing a first practical application of the exact
Eq. \eqref{eq_eff73},  here we will show that a similar approach for the design of new approximations to the
tensor xc kernel of TDCDFT is also possible. In fact, one expects to find a similar level of accuracy also
for current-current response functions.
By a first-order linearization of Eq. \eqref{eq_eff73}, where for the various response functions one uses the independent-particle versions, and taking for $\tilde{\Xi}$ the statically screened $W$, as usually done in BSE, one obtains:
\begin{multline}
\hat{f}_{xc,\alpha\beta}^{(2)}(1,2) = -\bigl[\hat{\chi}_0^{-1}\bigr]_{\alpha\lambda}(1,3) 
\contra_{\lambda\kappa} \left[ L_0(3,3',4,5)  W(4,5) \right. \\ \left. L_0(4,5,6,6') \right] \bigl[\hat{\chi}_0^{-1} \bigr]_{\kappa\beta}(6,2).
\label{eq_eff74a}
\end{multline}
More explicitly:
\begin{multline}
\hat{f}_{xc,\alpha\beta}^{(2)}(1,2) = -\bigl[\hat{\chi}_0^{-1}\bigr]_{\alpha\lambda}(1,3) \,\,  \\  
 \lim\limits_{\substack{3' \rightarrow 3 \\ 6' \rightarrow 6}}  \left[
 \frac{1}{2i}  (\nabla_{3_\lambda}-\nabla_{3'_\lambda})  L_0(3,3',4,5)   W(4,5) \right. \\ \left.
\frac{1}{2i}  (\nabla_{6_\kappa}-\nabla_{6'_\kappa}) L_0(4,5,6,6') \right] \,\,  \bigl[\hat{\chi}_0^{-1}\bigr]_{\kappa\beta}(6,2),
\label{eq_eff74}
\end{multline}
where  the presence of three-point current-density and density-current response functions, $\chi_{\mathbf{j}\rho}^3(3;4,5)$ and $\chi_{\rho\mathbf{j}}^3(4,5;6)$, becomes apparent. Hence the kernel \eqref{eq_eff74} can be also rewritten in a compact way as: 
\beq
\hat{f}_{xc}^{(2)} = - \hat{\chi}_0^{-1} \chi_{0,\mathbf{j}\rho}^3 W \chi_{0,\rho\mathbf{j}}^3 \hat{\chi}_0^{-1}.
\label{eq_eff75}
\eeq
This is a new approximation of the TDCDFT $\hat{f}_{xc}$ kernel that has to be understood as an orbital functional, hence an implicit functional of the current density. In this sense it is more flexible than the VK functional, which instead is an explicit functional of the current density.

The spatial derivatives that appear in  Eq. \eqref{eq_eff74} don't modify the structure of the poles of the response functions in the frequency domain.  Therefore, this approximation of the TDCDFT tensor xc kernel can benefit from the same cancellation of poles and zeroes of the response functions that has been shown to be essential for the Nanoquanta kernel of TDDFT \cite{francesco} (provided that the response functions entering Eq. \eqref{eq_eff74} are built with QP energies).  Since any scalar potential with a gauge transformation can be represented by a longitudinal vector potential, Eq. \eqref{eq_eff74} can be thought as a generalization of the Nanoquanta TDDFT $f_{xc}$ kernel to the calculation of the response to any kind of time-dependent external vector potential.

In general the relation between the tensor $\hat{f}^{(2)}_{xc}$ TDCDFT kernel and the scalar $f^{(2)}_{xc}$ TDDFT kernel is rather involved\cite{nazarov1}. Its first-order linearization in $\hat{f}^{(2)}_{xc}$ reads\cite{nazarov2}:
\beq
f^{(2)}_{xc} = - \frac{c}{\w^2} \chi_{0,\rho\rho}^{-1} \nabla \hat{\chi}_0 \hat{f}^{(2)}_{xc}  \hat{\chi}_0 \nabla \chi_{0,\rho\rho}^{-1}
\label{fxcfxc}
\eeq
This approximation is consistent with the linearization of Eq. \eqref{eq_eff73} that leads to \eqref{eq_eff75}.
Hence, by inserting the TDCDFT kernel $\hat{f}_{xc}^{(2)}$ \eqref{eq_eff75} in Eq. \eqref{fxcfxc} and using the density continuity equation\cite{foot1}, the TDDFT $f^{(2)}_{xc}$  kernel becomes:
\beq
f^{(2)}_{xc} = -\chi_{0,\rho\rho}^{-1} \chi_{0,\rho\rho}^{3} W \chi_{0,\rho\rho}^{3} \chi_{0,\rho\rho}^{-1},
\label{nanoquantafxc}
\eeq
which is, in a compact form, the Nanoquanta TDDFT kernel\cite{fabien,longrange,nanoquanta,stubner,ulf,SSE}. This equivalence, to the first order in $\hat{f}^{(2)}_{xc}$,  between the tensor kernel \eqref{eq_eff75} and the scalar kernel \eqref{nanoquantafxc} supports the validity of the new approximation for the TDCDFT kernel \eqref{eq_eff75}, thanks to the excellent results found using the Nanoquanta TDDFT kernel \eqref{nanoquantafxc}\cite{botti}. Moreover, this represents a further alternative derivation of the expression \eqref{nanoquantafxc} of the Nanoquanta kernel that exploits the possibility of mapping approximations developed in the framework of TDCDFT into scalar TDDFT kernels\cite{nazarov1,nazarov2,neepa3,lrvignale}.

In situations where the screening of the Coulomb interaction is ineffective, $W$ can be approximated with $v$ and the GW approximation reduces to Hartree-Fock.  In this case, \eqref{eq_eff74} reduces to an exact-exchange approximation for $\hat{f}_{xc}^{(2)}$ (which has been already worked out for the homogeneous electron gas in Ref. \cite{ulf}). On the other side, it has been demonstrated within the TDDFT framework that the lack of screening of the long-range contribution of the Coulomb interaction in the kernel implicitly overestimates both QP band gaps and excitonic effects, leading to pathologies  in optical spectra of semiconductors\cite{fabien2}. These pathologies can be cured by taking into account the screening of the Coulomb interaction, as done in Eq. \eqref{eq_eff74}.

For optical spectra of solids, relevant is the long-wavelength limit $\bfq \rightarrow 0$ of the $\bfG={\bfG}'=0$ element of the xc kernel written in reciprocal space: $\hat{f}_{xc,\alpha\beta}^{(2)}(\bfq+\bfG,\bfq+{\bfG}',\w)$, where $\bfG$ and ${\bfG}'$ are reciprocal lattice vectors. In this limit  the $\hat{f}_{xc}^{(2)}$ kernel \eqref{eq_eff74} in insulators becomes\cite{foot2} $\hat{\alpha} / \w^2 $. 
Moreover, the static $W$ is proportional to $1/\epsilon_{\infty}$, where $\epsilon_{\infty}$ is the static dielectric constant. 
This implies that for the $\bfq \rightarrow 0$  limit also $\hat{\alpha}$ is  proportional to $1/\epsilon_{\infty}$,
suggesting that in isotropic systems $\alpha$ could be used as a fitting parameter for the calculation of optical spectra in solids\cite{longrange,silvana}.
In this limit, the tensor kernel becomes completely local, contrary to the scalar TDDFT kernel that is ultranonlocal\cite{vignale2} and has a $1/q^2$ asymptotic behavior in the long-wavelength limit\cite{longrange,ggg}.

Here it is interesting also to note that, even though for the BSE kernel a static approximation for $W$ is adopted, the resulting $\hat{f}_{xc}^{(2)}$ kernel  in \eqref{eq_eff74} in general {\it is} naturally frequency dependent\cite{maitra}.
This is a consequence of the conversion of the spatial nonlocality into a frequency dependence\cite{SSE}, associated to the reduction of the number of degrees of freedom, when one passes from the four-point $\tilde{L}$ to the two-point $\tilde{\chi}$ and from the four-point $\tilde{\Xi}$ to the two-point $\hat{f}_{xc}$.

Since the validity of the new $\hat{f}_{xc}^{(2)}$ kernel \eqref{eq_eff74} is not confined to the weakly inhomogeneous limit, it could be for instance  used also to study dynamical corrections in the weak bias limit of the Landauer formula in molecular transport for nanoscale junctions\cite{landauer}.  In fact, so far the calculated corrections have been based on the VK functional, which is strictly valid only for slowly-varying densities in a high-frequency regime. Therefore an accurate estimate of these effects is still under debate\cite{landauer2}.
On the other side,  the renormalization of molecular electronic KS levels at metal-molecule interfaces\cite{renorm} in the present context is accounted for by the term $\hat{f}_{xc}^{(1)}$.

In conclusion, in the present work  we have derived an exact equation for the xc kernel $\hat{f}_{xc}$ of TDCDFT, which allows one to map well established working approximation of MBPT into the more efficient TDCDFT scheme, Due the lack of an appropriate approximate xc kernel, so far TDCDFT could not be used for calculating optical spectra in good agreement with experiments\cite{metano,berger,deboeij}.
So, in analogy with the successful results obtained in the TDDFT case \cite{botti}, we have suggested a practical approximation for $\hat{f}_{xc}$ and discussed its validity for the calculation  of electronic spectra of solids and nanosystems and dc conductances in molecular devices. 

We are grateful to Lucia Reining, Valerio Olevano and Ilya Tokatly for their comments and suggestions on many aspects of this work.
We acknowledge support from the EU's 7th framework program through the ETSF e-I3 infrastructure project  (Grant No. 211956), 
the Spanish MEC (FIS2007-65702-C02-01), ACI Promociona (ACI2009-1036), ‘Grupos Consolidados UPV/EHU del Gobierno Vasco’ (IT-319-07), and ETORTEK projects.

  \bibliographystyle{apsrev}

\begin{thebibliography}{99}
\bibitem{foot1} The continuity equation is strictly valid only if conserving approximations for the response functions are used\protect\cite{Strinati}.
\bibitem{foot2} In the present approximation and considering for all the quantities only the $\bfG={\bfG}'=0$ element, while $W$ is static and proportional to $1/q^2$, at low frequencies current-current response functions are proportional to $\w^2$ and density-current or current-density response functions are proportional to $\w$ and $q$.
\end{thebibliography}

  \end{document}